\documentclass{article}
\usepackage{arxiv}

\usepackage{pifont}

\usepackage{multirow}

\usepackage{booktabs}
\usepackage[ruled,vlined]{algorithm2e}
\usepackage{amsmath}
\usepackage{caption}
\usepackage{float}
\usepackage{graphicx}
\usepackage{subcaption}
\usepackage{natbib}
\usepackage{hyperref}
\usepackage{url}

\title{Deep Natural Language Processing for \\ LinkedIn Search}

\author{Weiwei Guo, Xiaowei Liu, Sida Wang, Michaeel Kazi, Zhiwei Wang,\\ {\bf Zhoutong Fu, Jun Jia, Liang Zhang, Huiji Gao, Bo Long}\\
         \texttt{\{wguo,xwli,sidwang,mikazi,zhiwwang,zfu,jjia,lizhang,hgao,blong\}@linkedin.com}\\
         LinkedIn, Mountain View, CA}

\begin{document}

\maketitle



\begin{abstract}
Many search systems work with large amounts of natural language data, {\it e.g.}, search queries, user profiles, and documents. Building a successful search system requires a thorough understanding of textual data semantics, where deep learning based natural language processing techniques (deep NLP) can be of great help.  In this paper, we introduce a comprehensive study for applying deep NLP techniques to five representative tasks in search systems: query intent prediction (classification), query tagging (sequential tagging), document ranking (ranking), query auto completion (language modeling), and query suggestion (sequence to sequence). We also introduce BERT pre-training as a sixth task that can be applied to many of the other tasks. Through the model design and experiments of the six tasks, readers can find answers to four important questions: (1) \textit{When is deep NLP helpful/not helpful in search systems?} (2) \textit{How to address latency challenges?} (3) \textit{How to ensure model robustness?}
This work builds on existing efforts of LinkedIn search, and is tested at scale on LinkedIn's commercial search engines. We believe our experiences can provide useful insights for the industry and research communities.

\end{abstract}


\maketitle

\section{Introduction}
\begin{figure}[ht]
\centering
  \includegraphics[scale=0.5]{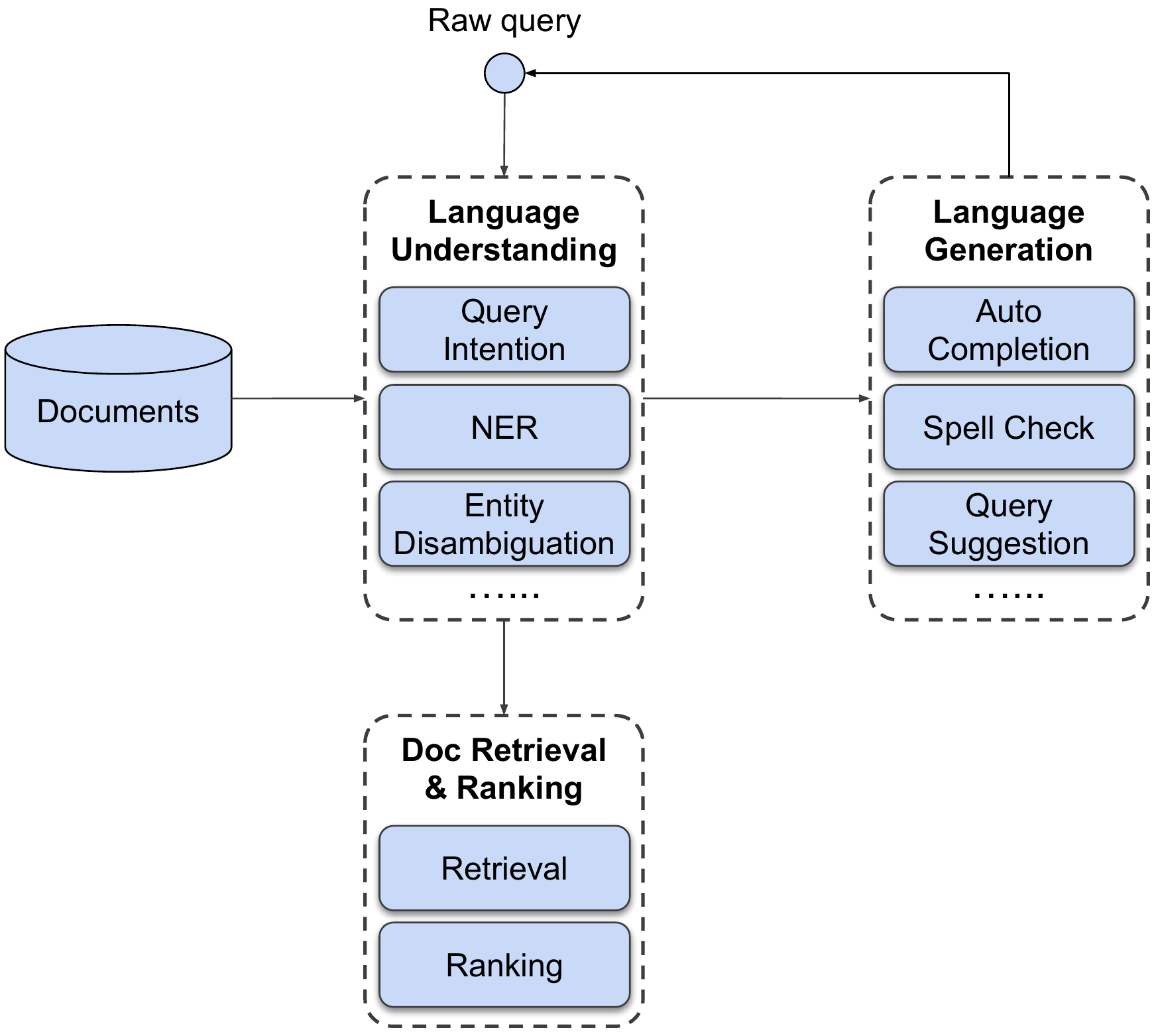}
  \caption{Overview of a search system.}
  \label{figure:overview}
\end{figure}

Search systems are widely used to help users retrieve information from large and often difficult to categorize data sources. Search engines are typically complicated ecosystems that contain many components. As shown in Figure \ref{figure:overview}, after a user issues a query, the language generation modules \citep{bar2011, li2006, cao2008} serve as search assistants to generate a better query. The language understanding modules \cite{kang2003query,guo2009,blanco2015} aim to extract semantic information from the query and documents, such as user intent and named entities. Finally, all the extracted information is used for document retrieval and ranking. 

A common part of these search components is that they all deal with large amounts of text data, such as queries, user profiles and documents.  Text data is sequential, and understanding such sequential information is a nontrivial task with traditional methods. For example, the systems need to handle (1) synonyms/similar concepts, {\it e.g.}, "software engineer" vs "programmer", (2) disambiguation, {\it e.g.}, "job opening" vs "restaurant opening", and (3) word importance weighting, {\it e.g.}, the important words in a query "looking for a research scientist job" would be "research scientist", and so on. Traditional methods highly rely on sparse features such as unigram features, which do not have strong generalization power to handle these cases.

On the other hand, deep learning has shown great success in NLP tasks \cite{lecun2015}, indicating its potential in search systems.  In addition to alleviating the language problems mentioned above, there are several other benefits of using deep learning, such as end-to-end training, automatic feature engineering, etc.

Developing deep NLP models for search systems requires considering three challenges of the complicated ecosystem of search engines \cite{Croft2010,mitra2018}. Firstly, serving \textbf{latency} constraints preclude complex models from being used in production. In addition, directly trained deep learning models can have \textbf{robustness} issues such as overfitting. The last challenge is \textbf{effectiveness}: often production models are very strong baselines that are trained on millions of data examples with many handcrafted features, and have been tweaked for years.

In this paper, we focus on developing practical solutions to tackle the three challenges, and share real world experiences.  As shown in Figure \ref{figure:tasks}, we have picked five representative search tasks that cover the classic NLP problems: classification, ranking, sequence tagging, language modeling, and sequence-to-sequence.  For each of the tasks, we investigate the unique challenges, provide practical deep NLP solutions, and analyze the offline/online experiments results. By providing a comprehensive study, we hope the readers can not only learn how to handle the different challenges in search systems, but also generalize and apply it to new tasks in other industry productions such as recommender systems.

The contribution of the paper is:
\begin{itemize}
    \item To our best knowledge, this is the first comprehensive study for applying deep NLP models in search productions.  Five tasks in search systems are selected, which cover most common NLP problems. For each task, we provide practical solutions, and report experiments in LinkedIn's commercial search engines.
    \item We go beyond the task boundary and summarize the observations and solutions into lessons.  We believe our experience would be a valuable resource for other vertical searches, as well as other applications such as recommender systems.
    \item We successfully deploy BERT to two real world productions: query intent and document ranking.

\end{itemize}

\section{Search Systems at LinkedIn}

\subsection{LinkedIn Search Systems Overview}
LinkedIn provides multiple vertical searches, each corresponding to a document type, e.g.,  \textit{people}, \textit{job}, \textit{company}, etc. In this paper, the experiments are conducted on 3 vertical searches (\textit{people search, job search, help center search}), and \textit{federated search}. Federated search retrieves the documents from all vertical searches and blends them into the same search result page. When users go to LinkedIn, federated search is the default.  People search is the most popular search engine, which retrieves member profiles; job search returns job posts for job seekers; help center search provides answers on how to use LinkedIn with a lot of natural language queries, \textit{e.g.}, "how to change my account password".

A typical search system is shown in Figure \ref{figure:overview}.  There are three main components: (1) \textbf{Language understanding} constructs important features for the other two components; (2) \textbf{Language generation} improves user experiences by suggesting queries that are more likely to lead to desirable search results.    (3) \textbf{Document retrieval \& ranking} produces the final results of search systems and presents them to users. 



\subsection{Characteristics of Vertical Search Data}
Search data is different from classic NLP task data, mainly from two aspects: data genre and training data size.  In classic NLP datasets \cite{pang2005,hu2004}, the data unit is one or several complete sentences with dozens of words in proper grammar.  In a search system, queries have several keywords without grammar, which introduces ambiguity.  Meanwhile, LinkedIn searches are vertical searches instead of web search such as Google. The vocabulary is mostly domain specific entities, \textit{e.g.}, people names, company, skills, etc.

In addition, the training data size is noteworthy.  Classic NLP datasets are human annotated, therefore the size is usually around tens of thousands of sentences, \textit{e.g.}, 11k training examples in the sentiment analysis dataset \cite{pang2005}.  However, search training data is usually derived from click-through data, hence contains millions of noisy training examples.

\subsection{Challenges of Deep NLP for Search}
There are several common challenges to applying deep NLP models to search systems. The first challenge is the online production \textbf{latency}.  Deep learning models are known for their large compute time.  Assuming the time complexity of a bag-of-words model is $O(n)$, where $n$ is the number of words in a sentence, then a LSTM model \cite{Hochreiter1997} has a time complexity of $O(nd^2)$, where $d$ is the number of dimensions used for word embeddings. In this paper, we analyze the unique latency challenge faced by each task, and provide multiple practical solutions to resolve or mitigate the latency issue.

The second challenge is \textbf{robustness}.  Deep learning models have many parameters, hence are more likely to overfit to the training data, and ignore the infrequent patterns.  For example, in query tagging, the LSTM models always recognize the query "linkedin facebook" as one company, since queries with two companies are rare. In addition, deep NLP models tend to over-generalize word semantics, \textit{e.g.}, in people search, professor profiles are matched to the query "student". In this paper, we show several tasks that could lead to overfitting, and provide practical solutions such as more careful training data creation to tackle it.

The third challenge is \textbf{effectiveness}.  The traditional production models are usually optimized for many iterations and are trained with millions of examples, hence hard to beat.  In general, we reuse the existing handcrafted features to alleviate the issue. We also analyze why deep learning models do not work well in some scenarios.

\begin{figure}[ht]
\centering
  \includegraphics[scale=0.5]{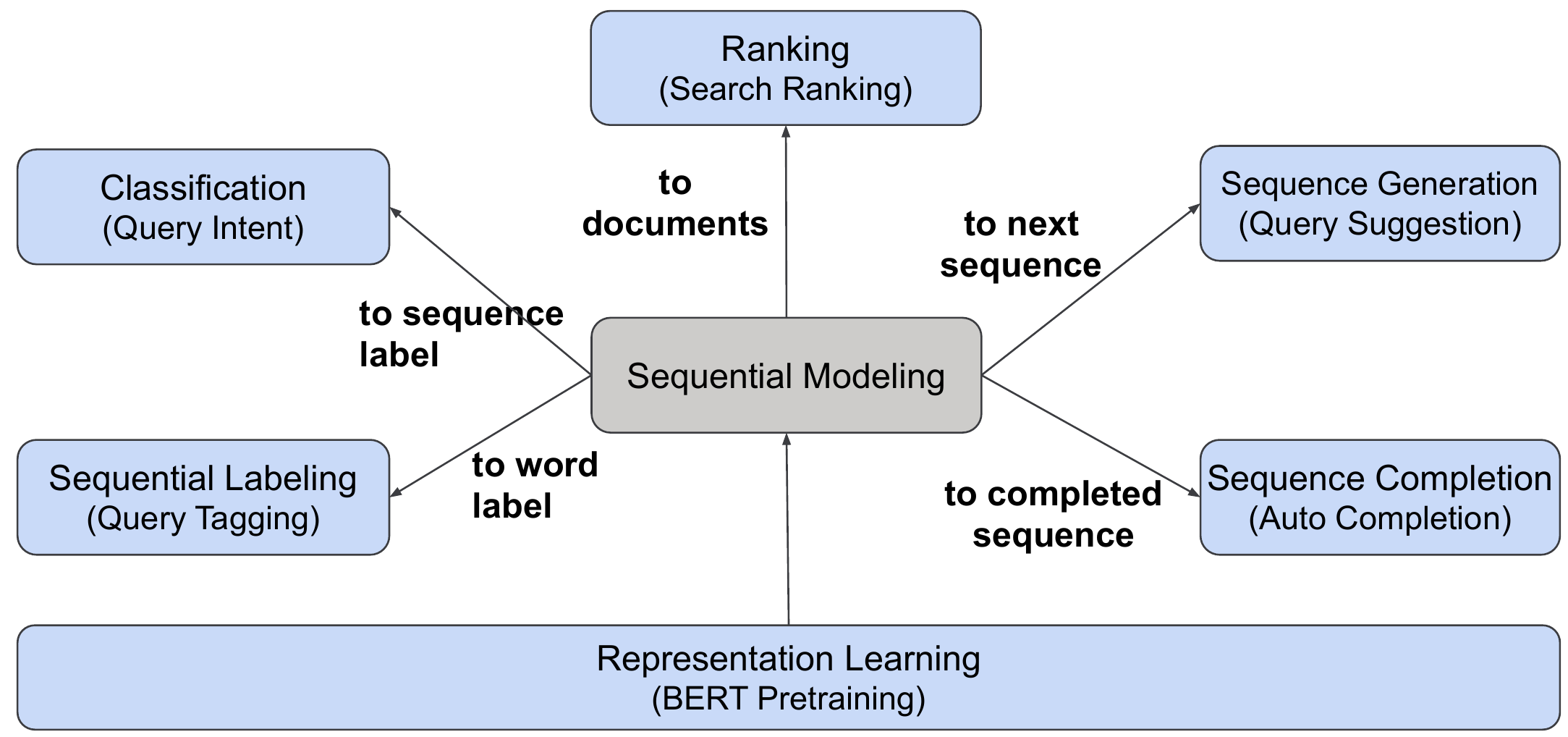}
  \caption{Deep NLP models for representative search tasks.}
  \label{figure:tasks}
\end{figure}

\subsection{Search Tasks}
The core of deep NLP is sequential modeling with CNN/\-LSTM/\-Transformer \cite{Lecun1995,Hochreiter1997,vaswani2017} networks to generate word or sequence embeddings. By adding specific loss functions on top of these embeddings, they can serve different NLP problems: predicting a sequence level label is a classification problem; predicting a label for each word is a sequential tagging problem, etc. In total there are five common NLP problems: classification, ranking, sequence tagging, language modeling, and sequence-to-sequence.
For each of the NLP tasks, we present a corresponding search task in Figure \ref{figure:tasks}.  These search tasks cover major components in search systems, ranging from language understanding/generation to document ranking. We also present a fundamental model pre-training task \cite{devlin2019} that can potentially benefit all tasks.

\section{Representative Search Tasks}
\label{section:five-task}
In this section, we introduce each specific task, outline the challenges, show how to overcome the challenges, and analyze the offline/online experiment results.

The experiments are conducted on the LinkedIn English market.  Offline results are reported on the test set. All reported online metrics are statistically significant with $p < 0.05$.  

\subsection{Query Intent Prediction}
\subsubsection{Introduction} 
Query intent prediction \cite{kang2003query,hu2009understanding} is an important product in modern search engines. Query intent is used in federated search to predict the probability of a user's intent towards seven search verticals: \textit{people}, \textit{job}, \textit{feed}, \textit{company}, \textit{group}, \textit{school}, \textit{event}. 
The predicted intent is an important feature leveraged by downstream tasks such as search result blending \cite{li2008learning} to rank higher the documents from relevant search verticals.

The challenge of this task is there are very few words in a query, hence it is hard to disambiguate words, such as "michael dell" (person names) vs "dell engineer jobs" (company). Deep NLP models can alleviate this issue, especially BERT which produces contextualized word embeddings \cite{devlin2019} (more details can be found in Section \ref{section:bert-pretraining}).

\subsubsection{Approach}
\begin{figure}[ht]
\centering
  \includegraphics[scale=0.5]{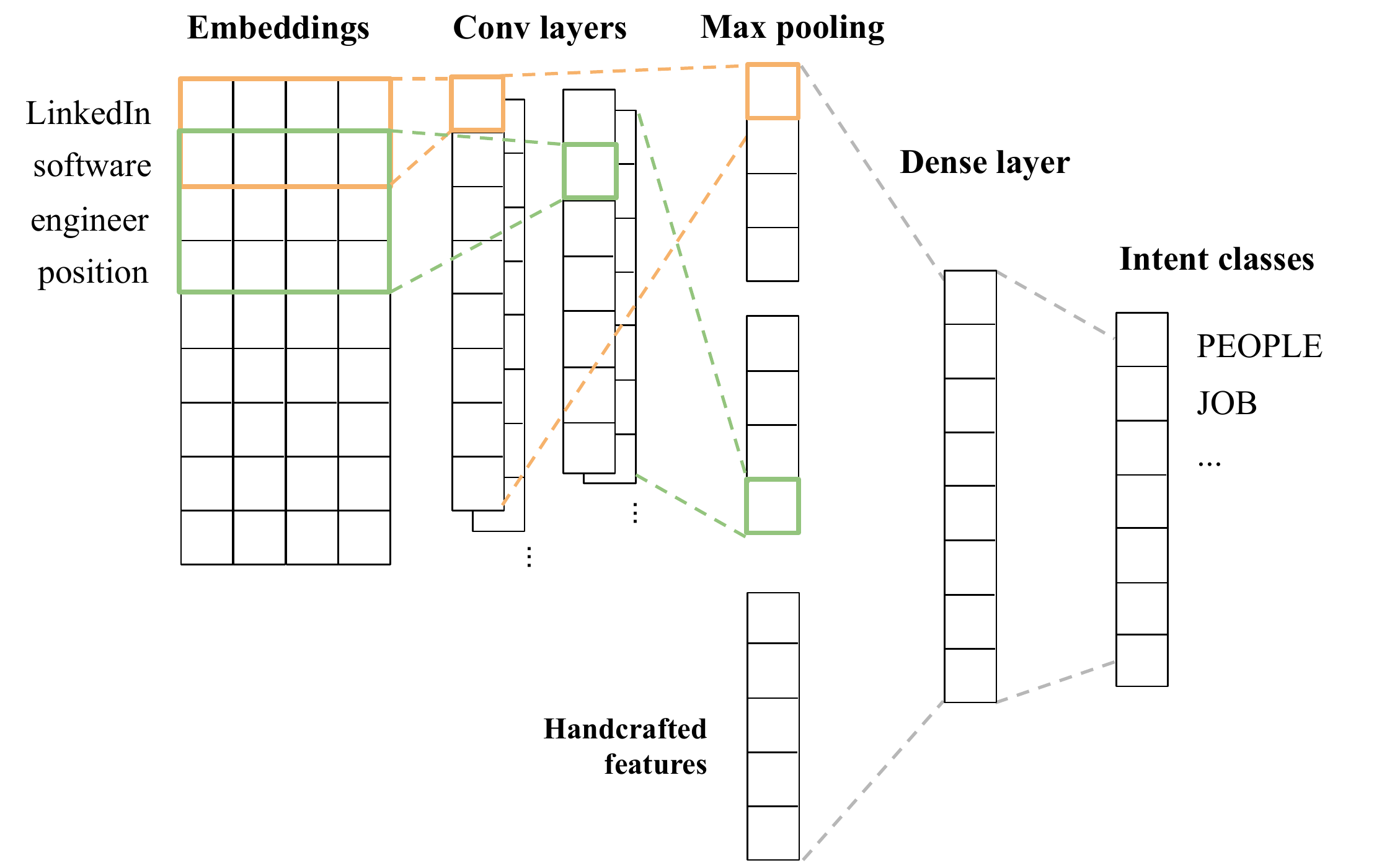}
  \caption{\small CNN based query intent prediction.}
  \label{figure:qim-cnn}
\end{figure}

The query intent prediction task is modeled as a multi-class classification problem. CNNs have achieved significant performance gains in text classification problems\cite{kalchbrenner2014convolutional,kim2014,hashemi2016}. In our approach (Figure \ref{figure:qim-cnn}), we combine the extracted text embedding with handcrafted features, and use a hidden layer to enable feature non-linearity. The existing handcrafted features are powerful. For example, the query tagger (Section \ref{section:qt}) feature can identify almost all people names accurately, including those out of the word embedding vocabulary. The member behavior features can enable personalization (e.g., whether a user clicks on job postings in a certain period of time).


\subsubsection{Experiments}
The label is inferred from click-through behaviors in the search log: if a user clicked on a document from one of the seven verticals, then the query is assigned the corresponding vertical label. We use 24M queries for training and 50k for dev and test. Besides the production baseline SCRF model, another baseline is bidirectional LSTM \cite{graves2013}. For the CNN and LSTM model, the vocabulary size is 100k with word embedding dimension as 64; word embedding is pre-trained with GloVe \cite{pennington2014};\footnote{In the five tasks except question suggestion, we always pre-train the word embedding on millions of training data with GloVe. In query tagger, we add additional queries and member profiles to enrich the corpus for pre-training.  We find word embedding pre-training always yields comparable or better relevance performance.} hidden layer dimension is 200 (Figure \ref{figure:qim-cnn}). In the CNN model, 128 filters of size 3 is used to capture word tri-grams. In LSTM, the hidden state size is 128.

\subsubsection{Results}
The baseline model is the production model, logistic regression on bag-of-words features and other handcrafted features. The offline relevance and latency performance are shown in Table \ref{table:offline-qim-cnn}. Both CNN and LSTM models outperform the production model, meaning that the features automatically extracted by CNN/LSTM can capture the query intents. For online experiments (Table \ref{table:qim-cnn-online}), we choose CNN instead of LSTM, since the relevance difference is small, but CNN is faster than LSTM. The online results show CNN increases the job documents click metrics.

\begin{table}
\centering
\caption{\small Offline comparison w.r.t. production baseline.}
\label{table:offline-qim-cnn}
\begin{tabular}{lcc}
\toprule
                  & \textbf{Accuracy}      & \textbf{P99 Latency} \\
\midrule
LR (baseline)          & - & -  \\
CNN w/o handcrafted features  & $+1.03\%$ & $+0.44$ms \\
CNN               & $+1.49\%$  & $+0.45$ms \\
LSTM              & $+1.61\%$  & $+0.96$ms \\
\bottomrule
\end{tabular}
\end{table}

\begin{table}
\centering
\caption{\small Online comparison for CNN based query intent prediction model vs production baseline. CTR@5 calculates the proportion of searches that received a click at top 5 items.
}
\label{table:qim-cnn-online}
\begin{tabular}{lc}
\toprule
    \textbf{Metrics}          &  \textbf{Percentage Lift} \\
\midrule
CTR@5 of job posts      & $+0.43\%$  \\
\bottomrule
\end{tabular}
\end{table}


\subsubsection{Related Work} The traditional methods use many handcrafted features, such as unigram, language model scores, lexicon matching features, etc \cite{arguello2009sources, cao2009context}. For Deep NLP models, offline experiments with CNN are conducted on 10k queries \cite{hashemi2016}. In contrast, our CNN model is trained on millions of queries with many handcrafted features (personalization, query tagger, etc). We also provided detailed analysis on latency and online impact on the vertical prediction task.

\subsection{Query Tagging}
\label{section:qt}
\subsubsection{Introduction} The goal of query tagging is to identify the named entities in queries. At LinkedIn, we are interested in 7 types of entities: \textit{first name, last name, company name, school name, geolocation, title, and skill}. After entities are identified, many important features can be constructed for downstream tasks such as query intent prediction or search ranking.

Query tagging is not a trivial task. For example, lexicon matching cannot solve the simple case such as "research scientist", because there are three entities matched: "research" as skill, "scientist" as title, "research scientist" as title.  In addition, for ambiguous queries, the query tagger should produce the most probable hypotheses, \textit{e.g.}, "vera wang" as a company rather than a person.

\subsubsection{Approach}
Query tagging is a named entity recognition task on query data. The production model uses three categories of features: character based, word based and lexicon based, as summarized in Table \ref{table:qt-ftr}. It is worth noting that we are able to extract powerful lexicon features leveraging large amount of user generated data, \textit{i.e.}, collecting the lexicon items from the corresponding fields of 600 million member profiles. Because of this, we choose semi-markov conditional random field (SCRF) \cite{sarawagi2005} as a baseline model, which can better exploit lexicon features than CRF \cite{lafferty2001}. 

The bidirectional LSTM-CRF architecture \cite{lample2016} proves to be a successful model on classic NLP datasets \cite{sang2003}.  We further extend it to bidirectional LSTM-SCRF.  Essentially, the deep part, Bidirectional LSTM, is used to replace the word-based features in Table \ref{table:qt-ftr}.




\begin{table}
\centering
\caption{\small Features used in SCRF based query tagger.}
\label{table:qt-ftr}
\begin{tabular}{ll}
\toprule
\textbf{Type} & \textbf{Description}\\
\midrule
char based & prefix/suffix features, such as "er", "ist"  \\
word based & word \\
& lemma \\
& brown cluster id \cite{brown1992} \\
& bigram with previous word, bigram with next word \\
lexicon & profile lexicon, collected from member profiles \\
 & clickthrough lexicon, collected from clickthrough data \\
\bottomrule
\end{tabular}
\end{table}

\subsubsection{Experiments}
\label{section:qt-exp}

\begin{table}
\centering
\caption{\small Query tagging results, measure by F1 score.}
\label{table:qt-res}
\begin{tabular}{lll}
\toprule
\textbf{Model} & \textbf{Hand-crafted Ftrs} & \textbf{F1}\\
\midrule
SCRF (baseline) & char/word/lexicon & - \\
CRF & char/word/lexicon & $-0.6\%$ \\
SCRF-nolex & char/word & $-6.1\%$ \\
\midrule
LSTM-SCRF & char/lexicon & $-0.3\%$ \\
LSTM-SCRF-all & char/word/lexicon & $-0.1\%$ \\
\bottomrule
\end{tabular}
\end{table}

Queries from LinkedIn federated search are collected and manually annotated. Meanwhile, a few thousand queries are generated to overcome the robustness problem (explained later). In total, we have 100k training queries, 5k dev and 5k test queries. For all models, the Adagrad optimizer \cite{duchi2011} is used with learning rate $10^{-3}$; batch size is 100; word embedding size is 50.  This is tuned on a dev set.

\subsubsection{Results} The traditional method SCRF achieves the best results in Table \ref{table:qt-res}. LSTM-SCRF has all the handcrafted features except word based features, however, it cannot outperform the SCRF baseline, and neither can LSTM-SCRF-all with all handcrafted features.  Due to no significant offline experiment gain, these models are not deployed online.

We believe the major reason is the strength of lexicon features, therefore LSTM does not help much. The SCRF-nolex (without lexicon based features) performance also indicates lexicon features are the most important. Meanwhile, looking at the data, we found most entities are already covered by the lexicons that are built on large scales of data. Other reasons could be due to the data genre: Queries are much shorter than natural language sentences. Therefore, LSTM's ability to extract long distance dependencies is not helpful in this task.

\subsubsection{Related Work}
The early works use CRF/SCRF to extract entities in queries \cite{li2009extracting,guo2008unified}. SCRF \cite{sarawagi2005} yields stronger performance superior by better exploiting lexicon features.  The deep NLP models are mainly designed for natural language sentences with LSTM-CRF \cite{lample2016,Ma:16}, where LSTM is able to generate more powerful features by summarizing the long distance dependency. In this paper, we presented a strong production baseline SCRF with lexicons constructed from large scale datasets, and analyzed why deep NLP models failed to improve the relevance performance.

\subsection{Document Ranking}
\subsubsection{Introduction}
Ranked documents are the final results presented to users.  Given a query, a searcher profile and a set of retrieved documents, the goal is to assign a relevance score to each document and generate the ranking.

Latency is the biggest challenge for this task. Although it has the same time complexity as query intent/tagging tasks, the data unit of document ranking task is a set of documents, therefore the absolute time is not affordable.  The other challenge comes from effectiveness.  The production model is optimized in many iterations, with many strong handcrafted features.

In this paper, the experiments are conducted on people search and help center search.  Table \ref{table:ranking-data} shows the statistics of two searches. 

\subsubsection{Approach}
The production model uses XGBoost \cite{chen2016} for training, which works well with large scale training data and is effective in modeling feature non-linearity. In people search, many non-text features are used, such as personalized features based on social network structures and past user behaviors, and document popularity features based on search log statistics. In addition, millions of training examples are available for people search.  

\begin{table}
\centering
\footnotesize
  \caption{\small Statistics of two document ranking datasets.}
  \label{table:ranking-data}
  \begin{tabular}{lp{3cm}p{3cm}}
    \toprule
     & \textbf{People Search} & \textbf{Help Center} \\
    \midrule
    \# docs & 600M & 2,700\\
    \# training data (queries) & 5M & 340,000\\
  \bottomrule
\end{tabular}
\end{table}

\begin{figure}[ht]
\centering
  \includegraphics[scale=0.7]{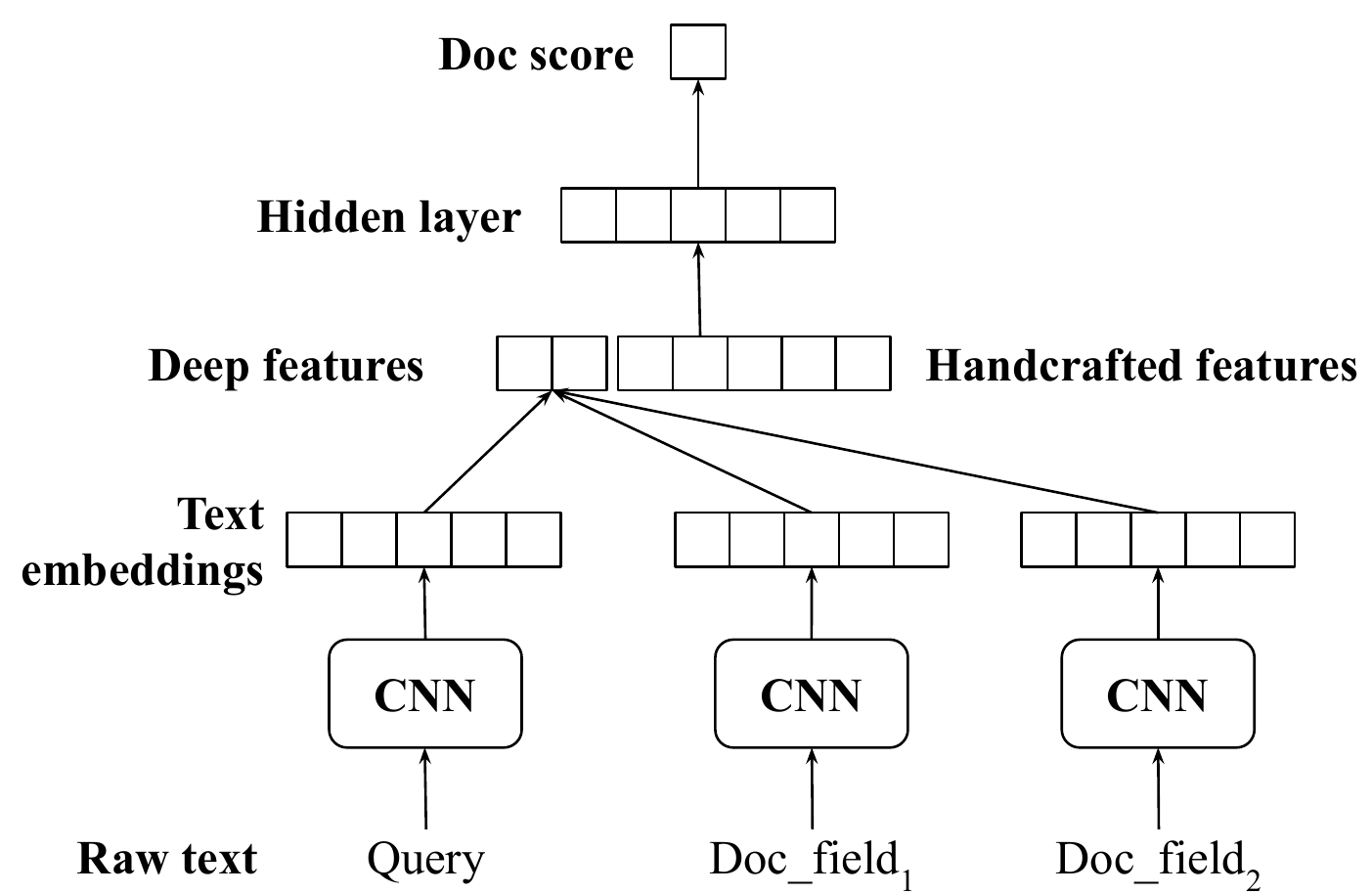}
  \caption{Model architecture for ranking model (The learning-to-rank layer is not included).}
  \label{figure:ranking-model}
\end{figure}

One benefit of a deep learning approach is that we can easily combine the deep NLP based semantic matching with other techniques that prove to be effective. Figure \ref{figure:ranking-model} shows the architecture.

In Figure \ref{figure:ranking-model}, the input is a query, multiple document fields, and a hand crafted feature vector.  After text embeddings are extracted, cosine similarity is computed for each query/document field embedding. The cosine similarity is combined with hand crafted features, and a Multi-Layer Perception \cite{pal1992} layer is used with a hidden layer, followed by a learning-to-rank layer. In general, we extend the previous work \cite{Huang2013,Zamani2018}, by combining hand crafted features with cosine similarities and adding a hidden layer. 

\noindent\textbf{Online Production Deployment}
As mentioned before, latency is a major issue for ranking tasks.  To reduce latency, we use two different online deployment strategies, based on the particular production environment. For the help center task, since there are only thousands of documents in total, we pre-compute the document embeddings.  By doing this, the computation is reduced to query embedding extraction and cosine similarity between query and document embeddings.

Pre-computing the document embeddings requires nontrivial infrastructure changes for people search, where there are 600 million documents (member profiles). It needs a lot of space to store the embeddings, as well as complicated designs to keep them fresh. 
In the P99 case in people search, there are thousands of retrieved documents on a searcher machine.\footnote{there are many searcher machines, each responsible for a part of the 600m profiles} Our solution is a two-pass ranking strategy: firstly apply a lightweight model without deep learning modules (MLP layer only), and choose the top hundreds of documents to send to deep models for reranking. After applying this change, the P99 latency is significantly reduced.

\subsubsection{Experiments}

\begin{table}
\centering
  \caption{\small Offline document ranking results (NDCG@10).}
  \label{table:ranking-offline}
  \begin{tabular}{lccc}
    \toprule
     \textbf{Models} & \textbf{People Search} & \textbf{Help Center}\\
    \midrule
    XGBoost (baseline) & - & - \\
    CNN-ranking &  $+3.02\%$ & $+11.56\%$\\
    CNN-ranking {\footnotesize w/o handcrafted features} &  $-4.52\%$ & $+11.07\%$\\
  \bottomrule
\end{tabular}
\end{table}

\begin{table}
\centering
\caption{\small Online document ranking results on two searches. "Sat click" is a people search metric: number of satisfactory searches that include (1) connecting, messaging or following a profile (2) viewing a profile with dwelling time > 5 seconds. "Happy path rate" is a help center search metric: proportion of users who searched and clicked a document without using help center search again in that day.}
\label{table:ranking-online}
\begin{tabular}{llc}
\toprule
\textbf{Search} & \textbf{Metrics} & \textbf{Percentage lift} \\
\midrule
\multirow{2}{*}{People Search} & CTR@5 & $+1.13\%$\\
 & Sat Click & $+0.76\%$\\
\midrule
Help Center & Happy Path Rate & $+15.0\%$\\
\bottomrule      
\end{tabular}
\end{table}

The experiments are conducted on two searches (Table \ref{table:ranking-data}). On average each query has around 10 documents. Dev and test size is 50k. For all models that apply, Adam optimizer \cite{kingma2015} is used with learning rate $10^{-3}$; batch size is 256; word embedding size is 64.

\subsubsection{Results} We report offline and online results in Table \ref{table:ranking-offline} and \ref{table:ranking-online}\footnote{The online setting is slightly different in help center search. Since there are only thousands of documents in total, the production model will score all the documents.  Therefore, for cnn-ranking we also adopt this setting.}, respectively. In general, the offline and online performance are consistent. The baseline is the production model trained with xgboost.
Our model \textit{CNN-ranking} significantly outperforms the baseline by a large margin.

To understand the impact of handcrafted features, we perform another experiment using CNN-ranking without any these features.  It turns out the handcrafted features in people search are more powerful than the help center, largely because people search contains a lot of social network features and document clickthrough features.

By comparing CNN-ranking vs baseline across the two searches, it is interesting to see that the impact of deep NLP models is significantly larger in the help center setting.  This is again caused by the data genre. In help center, the queries and documents are mostly natural language sentences, such as a query "how to hide my profile updates" to document "Sharing Profile Changes with Your Network", where CNNs can capture the semantic meaning. In people search, it is more important to perform exact matching, for example, query "facebook" should not return member profiles who work at "linkedin".

Finally, we present the latency of different deployment strategies in Table \ref{table:ranking-latency}.  For people search, we do not observe significant difference in terms of online relevance metrics between two pass ranking and all-decoding. We manually checked the ranking scores, and found that the documents discarded in the first phrase ranking usually have very low scores.

\begin{table}
\centering
\footnotesize
  \caption{\small{P99 latency in document ranking.}}
  \label{table:ranking-latency}
  \begin{tabular}{lccc}
    \toprule
    \textbf{Deployment Strategy} & \textbf{Search} & \textbf{\#Docs} & \textbf{@P99 latency}\\
    \midrule
    Two pass ranking & People search & 100-999 & +21ms\\
    All-decoding & & 1000-9999 & +55ms \\
    \midrule
    Document pre-computing & Help center & 1000-9999 & +25ms\\
  \bottomrule
\end{tabular}
\end{table}

\subsubsection{Related Work}
There are many existing works on deep NLP models for search ranking \cite{Huang2013,Shen:14,Palangi:16,guo2016,xiong2017,dai2018}. Our model adopted many designs in previous work to achieve balance between efficiency and effectiveness: multiple document fields \cite{Zamani2018} to improve document understanding, text-level interaction instead of word-level interaction \cite{Huang2013}, etc. In addition, we also show that combining existing handcrafted features with deep features in an MLP layer to optimize relevance performance. For production deployment, while the previous work focus on the embedding precomputing \cite{ramanath2018, yin2016, Grbovic2018}, we demonstrate that two pass ranking can work well with significantly less infra changes.

\subsection{Query auto completion}
\subsubsection{Introduction}
Query auto completion \cite{bar2011} is a language generation task.  The input is a prefix typed by a user, and the goal is to return a list of completed queries that match the user's intent. As a search assistance component, it improves user experience in two aspects: (1) Saves user keystrokes and returns search results in less time. (2) More importantly, it guides users to better queries and search results, \textit{e.g.}, for a prefix \textit{sof}, the query \textit{software engineer} is considered better than \textit{software developer}, since the former is a more common job title that leads to better recall.

Query auto completion has a strict latency requirement, since the model needs to return results for each keystroke.

\subsubsection{Approach}
\label{section:qac-prod}
The traditional auto completion system has two separate steps: candidate generation and candidate ranking.  The \textit{candidate generation} performs a lookup from a query prefix to a completed query, which is extremely efficient. This is done by memorizing each query prefix to the set of all possible queries that have been seen in the search log. For unseen query prefixes, a heuristic is used to generate the candidates \cite{mitra2015}. The \textit{candidate ranking} component is to rank the completed queries with frequency based counting or ML models like XGBoost. Several features are constructed for the completed queries, with the most effective feature being the query frequency. In summary, both candidate generation and candidate rank stages are lightweight and can be finished within several milliseconds.

Since query auto completion is a language generation task, an ideal model is neural language modeling \cite{Mikolov2010} with beam search decoding \cite{park2017}.  This neural language modeling approach achieves impressive relevance results, but at the cost of latency.  During beam search decoding, generation and ranking are performed at the same time for many iterations (one iteration for one generated token); while in traditional methods, there is only one generation and one ranking step. According to reported previous work \cite{park2017}, relevance performance can be increased by $10\%$, while latency can rise to over 1 second \cite{Wang2018}.

\begin{figure}[ht]
\centering
  \includegraphics[scale=0.7]{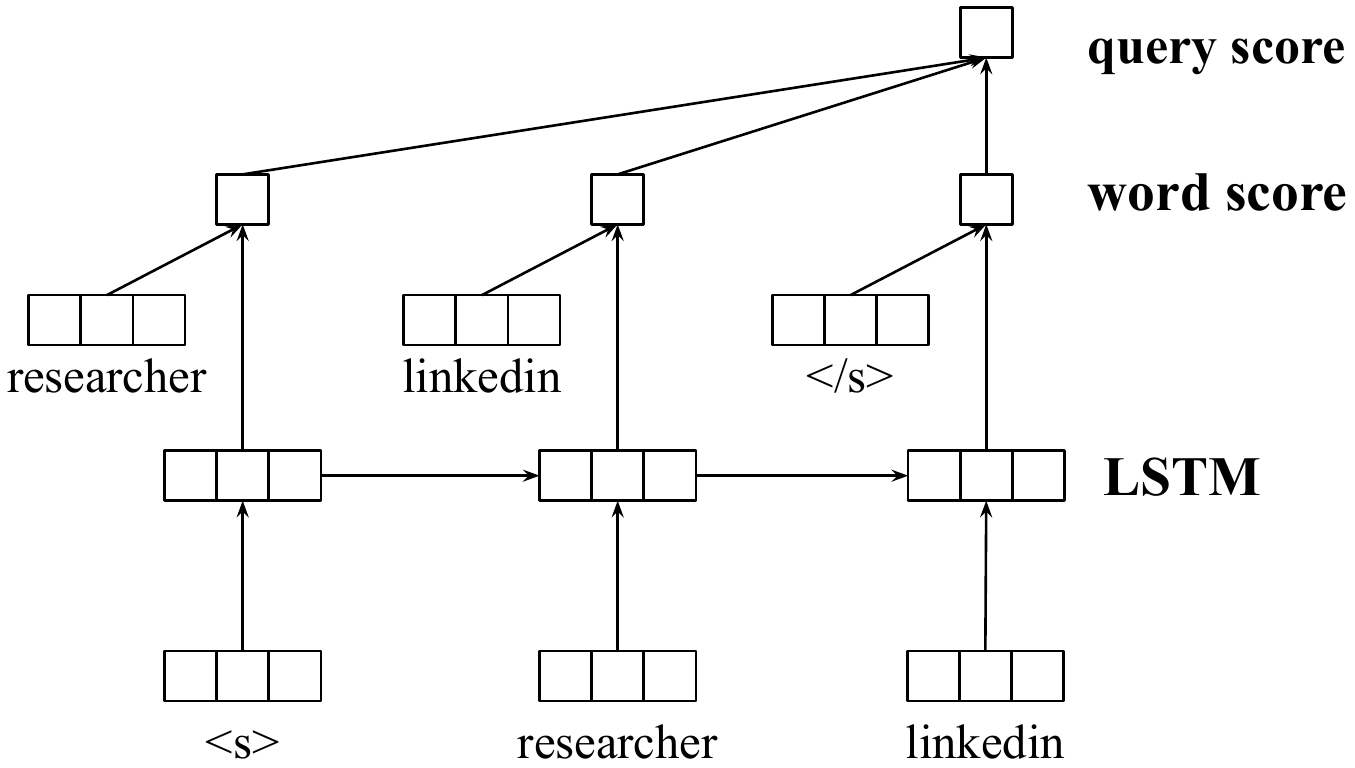}
  \caption{Model architecture for query auto completion. The learning-to-rank layer is not included.}
  \label{figure:qac-model}
\end{figure}

To reduce latency, instead of a end-to-end neural network approach, we apply deep learning only for the ranking component (Figure \ref{figure:qac-model}).   A LSTM based language model is used to assign a score for each candidate. We notice that the majority of time is spent on the computing the normalization constant of the word probability, since it sums over the entire vocabulary:
\begin{equation}
\log{P(w_i|h_i)}=\log{\frac{\exp(v_i^{\top}h_i)}{\sum_j{\exp(v_j^{\top}h_i)}}}=v_i^{\top}h_i-\log{\sum_j{\exp(v_j^{\top}h_i)}}
\end{equation}
Therefore, following the unnormalized language model approach \cite{sethy2015}, we approximate the denominator by $\log{\sum_j{\exp(v_j^{\top}h_i)}} = b$, where $b$ is another parameter to estimate.  The computation time reduction is summarized in Table \ref{table:qac-latency}.

\begin{table}
\centering
  \caption{\small P99 Latency in query auto completion task. Each prefix has 100 candidates to rank.}
  \label{table:qac-latency}
  \begin{tabular}{lc}
    \toprule
    \textbf{Models} & \textbf{Latency} \\
    \midrule
    Baseline & - \\
    Language Model (ranking only) & +61ms \\
    Unnormalized LM (ranking only) & +9ms \\
  \bottomrule
\end{tabular}
\end{table}

\subsubsection{Experiments}
\begin{table}
\centering
\small
  \caption{\small Offline relevance performance in query auto completion task, measured by mean reciprocal rank (MRR@10 \cite{craswell2009}).}
  \label{table:qac-offline}
  \begin{tabular}{lccc}
    \toprule
    Ranking Models & All & Seen prefix & Unseen prefix\\
    \midrule
    baseline  & -	& - & - \\
    Unnormalized LM & $+3.2\%$ & $+0.06\%$ & $+6.0\%$ \\
    CLSM \cite{mitra2015} & $+2.1\%$ & $+0.06\%$ & $+3.9\%$\\
    \bottomrule
  \end{tabular}
\end{table}

\begin{table}
\centering
  \caption{\small Online metrics of query auto completion at job search. "Job Applies" is the number of job posts applied from search. "Job Views" is the number of job posts clicked from search.s}
  \label{table:qac-online}
  \begin{tabular}{lc}
    \toprule
    Metric & Percentage Lift\\
    \midrule
    Job Applies & +1.45\% \\
    Job Views & +0.43\% \\
  \bottomrule
  \end{tabular}
\end{table}

The model is applied on job search query auto completion.  We follow the experiment setting and the train/dev/test splitting schema defined in \cite{mitra2015}.
The LSTM has one layer with a 100 dimension hidden state. Adagrad is used with learning rate $10^{-3}$. The vocabulary is 100k words with 100 dimensional embedding.

\subsubsection{Results} XGBoost is used as the product baseline model for candidate ranking component. Table \ref{table:qac-offline} and \ref{table:qac-online} show the offline/online results, respectively. All deep learning models outperforms the traditional methods by a large margin.  More interestingly, the gain is mostly from unseen query prefixes.  For seen prefixes, the frequency based methods can do a fairly good job.
We also compare with CLSM model \cite{mitra2015}. It formulates the problem as ranking the completed suffix given a prefix, \textit{i.e.}, the cosine similarity between the prefix and suffix is the ranking score. It has worse results than language modeling based methods, since the task nature is to rank the likelihood of completed queries, while CLSM focuses on extracting ngram patterns. 


\subsubsection{Related Work}
Traditional query auto completion systems use a two-step framework, candidate generation and ranking \cite{Cai:16}, similar to the production baseline described in Section \ref{section:qac-prod}. The deep NLP models in previous work perform candidate generation and ranking at the same time via beam search \cite{park2017}. This neural beam search framework is followed by adding personalization \cite{Jaech2018,fiorini2018,Jiang:18} and time information \cite{fiorini2018}. While impressive relevance performance is achieved, the model latency does not meet industry requirement. In this paper, we proposed a novel approach to effectively model the query context and reduce the latency.

\subsection{Query Suggestion}
\subsubsection{Introduction}
Query suggestion \cite{fonseca2005,cao2008} is another essential part of our search experience. Many search engines offer such a function, \textit{e.g.},  Google's "Searches related to ...", and LinkedIn's "People also search for", to assist users to seek relevant information.

\subsubsection{Approach}
The production baseline is based on frequency counting of search queries. It collects query pairs, and for each input query, sorts the suggestions by the query pair frequency. Heuristics are used to make sure the query pairs are semantically related: (1) The query pairs must be in the same session, where sessions are defined by queries separated by no more than 10 minutes; and either (2) two queries must share one common word or (3) the two queries co-occur for several distinct users.
 
We formulate the problem as machine translation in sequence-to-sequence (seq2seq) framework~\cite{Sutskever2014}, where the main benefit is to generalize to infrequent and unseen queries. We find the deep learning model can overfit to reformulation pairs if not careful handled.  When training data contain query \textit{generalization}, e.g. "research scientist --> scientist", the trained model will degrade to only delete words, to produce low perplexity suggestions. We handle this by removing these types of examples from the training data.

The seq2seq model has a large latency, which can be an issue in production. In this case, we serve our model in parallel with search result ranking, which gives us plenty of time (more than 100ms) to run the seq2seq model.



\subsubsection{Experiments}
The seq2seq model is tested in federated search. Training/dev/test data size is 300m/50k/50k query pairs. For the seq2seq model, we use a small model to reduce latency (dimension in LSTM is 100 units, 2 layers deep). Stochastic gradient descent is used with a learning rate of 1.0.

\subsubsection{Results} In our offline experiments in Table \ref{table:qs-offline}, we measure both the relative lift of MRR@10, and the coverage. The MRR of the model is computed based on the position of the gold standard query in the top 10 list, if it appears at all. The coverage indicates whether the model is able to produce any suggestion at all from the input. The MRR improves significantly relative to the baseline by using a deep learning method. The coverage of the deep learning method is trivially 100\%, although in practice it is lower due to filtering of unknown words or the rare blacklisted query. Overall, we can see that query suggestion is an area where deep learning provides great impact over traditional methods.

Online experiments (Table \ref{table:qs-online}) display evidence of an enhanced user experience, particularly on finding jobs and overall proportion of successful LinkedIn use.

\begin{table}
\centering
\caption{\small Offline experiments for query suggestion task.}
\label{table:qs-offline}
\begin{tabular}{lcc}
\toprule
\textbf{Model} & \textbf{MRR@10} & \textbf{Coverage} \\
\midrule
Frequency baseline & - & 67.3\% \\
Seq2Seq & +11.1\% & 100\% \\
\bottomrule
\end{tabular}
\end{table}

\begin{table}
\centering
\caption{\small Query suggestion online results.}
\label{table:qs-online}
\begin{tabular}{lc}
\toprule
Metric & Percentage Lift  \\
\midrule
Job views & $+0.6\%$ \\
Job apply & $+1.0\%$ \\
\bottomrule
\end{tabular}
\end{table}


\subsubsection{Related Work}
Traditional approaches to this problem are frequency based counting of search queries, or collaborative filtering \cite{paine2007recommending}, both of which only work on previously seen searches. He et al.\ \cite{He2016} applied seq2seq to rewrite queries into document titles, which is later used as features for search ranking. For query suggestions, multiple session models have been proposed \cite{Dehghani:2017,Ren2018} by exploiting the other queries in a search session. However, these works focus on offline results. In this paper, we outlined a practical solution for the related search production, from query reformulation collection to model training and online deployment. Our approach is applicable to any search engine with search logs.

\subsection{BERT Pre-training}
\label{section:bert-pretraining}
BERT \cite{devlin2019} is a pre-trained language representation model that is proven to be beneficial for many NLP tasks. In this paper, our goal is to pre-train a BERT model on LinkedIn data (hence the name LiBERT).  The advantage of a LinkedIn specific BERT model is: (1) Better relevance for domain specific tasks. In Google's pre-trained model, "linkedin" and many other companies are not in the vocabulary. (2) Smaller model structure for the ease of deployment. Our model is 6 layer with 34m parameters, compared to 12 layers and 110M parameters in Google's BERT-Base model.

\subsubsection{Model Pre-training}
The original BERT models were pre-trained on Wikipedia and BooksCorpus data, which are widely used datasets in language modeling. The text data in LinkedIn search systems are of a different genre. For pre-training the LiBERT models, we extracted data from different domains at LinkedIn.  Table \ref{table:bert-data} shows the data source of the pre-training data as well as the statistics. The large amount of text data from these domains should cover most NLP use cases in search products and are general enough for downstream fine-tuning tasks.

The major challenge of productionizing BERT is the latency.  In this paper, we train a smaller LiBERT on LinkedIn data to minimize the latency.  We experimented with the original BERT implementation \cite{devlin2019}, as well as our more light-weight LiBERT architecture (LiBERT). The model specification is 6 transformer layers (L=6), 12 hidden size (H=512), 8 attention heads (A=8), and 64 positional embeddings (P=64). A much smaller number of positional embeddings is used, due to that the length of search queries are much shorter than the sentences in classic NLP tasks.  The resulting model has 1/3 as many parameters as BERT-Base.






\begin{table}
\centering
\caption{\small Corpus used for LiBERT pre-training.}
\label{table:bert-data}
\begin{tabular}{llc}
\toprule
\textbf{Corpus} & \textbf{Description} & \textbf{\#Words}\\
\midrule
Search          &   Search queries      &   890M    \\
Member profiles &   Headlines, summaries, positions    &   728M    \\
Job listings     &   Job titles and descriptions &   604M    \\
Ads             &   Ads titles and descriptions &   637M    \\
\bottomrule
\end{tabular}
\end{table}

\subsubsection{Experiments on Fine-tuning Tasks}
Fine-tuning experiments are conducted on two tasks: help center search ranking and query intent classification. For help center ranking, the deployment strategy is document embedding pre-computing. The CNN model in the classification and ranking task (Figure \ref{figure:qim-cnn} and \ref{figure:ranking-model}) is replaced by our LiBERT model. Since queries are short, we only keep the first 16 words in queries, to further reduce P99 computation. 

In Table \ref{table:bert-offline}, the LiBERT model achieves comparable performance to BERT-Base in both tasks, even though a simpler architecture is used. This is due to the in-domain pre-training data that is more relevant to the task.  Similar trend of performance improvements are observed in online experiments, as shown in Table \ref{table:bert-online}. 
We also reported the P99 offline latency.  In the help center search task, recall that document pre-computing is used, and all 2,700 document embeddings are directly compared to the queries.   As shown in Table \ref{table:bert-latency}, LiBERT can significantly reduce the computation time, compared to BERT-Base. Accordingly, we deployed the LiBERT based models to production.


\begin{table}[]
\centering
\footnotesize
\caption{\small LiBERT fine-tuning offline performance comparison on query intent and help center ranking tasks.}
\label{table:bert-offline}
\begin{tabular}{lcccccc}
\toprule
\textbf{Model} & \multicolumn{4}{c}{\textbf{BERT HParams}} & \textbf{Query intent} & \textbf{Help center} \\
 & $\#L$ & $\#H$ & $\#A$ & $\#P$ & Accuracy & NDCG@10 \\
\midrule
CNN & - & - & - & - & - & - \\
BERT-Base & $12$ & $768$ & $12$ & $512$ &  $+2.89\%$ & $+2.15\%$\\
LiBERT    & $6$  & $512$ & $8$ & $64$ &  $+3.28\%$ & $+2.13\%$ \\
\bottomrule                            
\end{tabular}
\end{table}

\begin{table}[]
\centering
\caption{\small Online experiments of LiBERT over the baseline model (CNN) in query intent prediction and help center ranking tasks.}
\label{table:bert-online}
\begin{tabular}{llc}
\toprule
\textbf{Task} & \textbf{Metrics} & \textbf{Percentage lift} \\
\midrule
Query intent & CTR@5 &  $+0.17\%$\\
 & SAT click & $+1.36\%$\\
\midrule
Help center & Happy Path Rate &  $+11.3\%$\\
\bottomrule      
\end{tabular}
\end{table}

\begin{table}[]
\centering
\caption{\small Offline P99 latency of LiBERT on query intent classification and help center ranking.}
\label{table:bert-latency}
\begin{tabular}{lcc}
\toprule
\textbf{Model} & \textbf{Classification} & \textbf{Ranking} \\
\midrule
CNN & +0.5ms & +25ms\\
BERT-Base & +53ms & +65ms \\
LiBERT & +15ms & +44ms \\
\bottomrule      
\end{tabular}
\end{table}

\section{Lessons Learned}
Our previous section focuses on task specific challenges and solutions for each individual task.  In this section, we go beyond the task boundary and generalize the interesting observations into lessons.  We believe these lessons are not limited to search systems, but also can be applied to other areas such as recommender systems.

\subsection{When is Deep NLP Helpful?}
In general, the deep NLP models achieve better relevance performance than traditional methods in search tasks, and are particularly powerful in the following scenarios:
\begin{itemize}
\item \textbf{Language generation tasks}. Based on the offline experiments, query suggestion (Table \ref{table:qs-offline}) benefits most from deep NLP modeling. This is determined by the nature of language generation tasks. 
In the seq2seq framework, the hidden state summarizes the context well, and the decoder can produce related queries for any input query.\\ 
\item \textbf{Data with rich paraphrasing}.  In document ranking, the improvement brought by using CNN is larger in help center than in people search (Table \ref{table:ranking-offline} and \ref{table:ranking-online}), since help center has a lot of natural language queries while people search queries are combination of keywords.
\end{itemize}

\subsection{When is Deep NLP not Helpful?}
There are two cases in search systems where no gain is observed by applying deep NLP models: (1) query tagging (Table \ref{table:qt-res}); (2) seen prefixes in query auto completion (Table \ref{table:qac-offline}). 

There are mainly two reasons for this.  The first reason is for the particular task, the handcrafted features are powerful enough to solve most of the problems.  In query tagging, there is a lexicon built on 700 million user profiles, therefore the lexicons can give a pretty accurate estimation of the entity label. Similarly, for seen prefixes of query auto completion, the frequencies of prefixes to completed queries is collected on million of entries in the search logs, which makes them reliable.

The second reason is the data genre.  Firstly, query tagging datasets contain a lot of people names (over 50\%) where word embedding do not help much. In fact, except for query intent and query tagging, the other three tasks work on datasets without people names.  Secondly, it lacks language variation, similar to what we mentioned in the last section.  It is interesting that in classic NLP data sets, e.g. CoNLL'03 \cite{sang2003}, the LSTM-CRF \cite{lample2016} outperforms CRF by a large margin. In the NLP data set, each data unit is a complete sentence, which provides context clues. For example, the headline "Jordan expels Iraqi diplomat" makes it obvious what kind of entity "Jordan" is, but it would be rare to see this in search data.

\subsection{Latency is the Biggest Challenge}
We found latency to be the biggest challenge of applying deep NLP models to search productions. We summarize as below the practical solutions to reduce latency:
Dense matrix multiplication and evaluating softmax are costly, and these are often performed on every word in a text. 
\begin{itemize}
    \item \textbf{Algorithm redesign}. In query auto completion, instead of a fully neural network approach, we only apply deep learning for candidate ranking. In addition, the unnormalized language model is applied to further reduce latency.
    \item \textbf{Parallel computing}.  In query suggestion, the seq2seq model runs in parallel with search ranking, which leave more than 100ms buffer for seq2seq model.
    \item \textbf{Embedding pre-computing}. In help center search ranking, the document embeddings are pre-computed, leaving only query processing to be computed at run-time.
    \item \textbf{Two pass ranking}.  In people search ranking, a lightweight model is applied to handle all documents, then a deep model reranks the top ones.  Compared to embedding pre-computing, two pass ranking has less of an infrastructure requirement.

\end{itemize}

It is also worth noting that due to the nature of the production setting, computation time may not be a blocker.  In query intent, the BERT model is computationally heavy, but it only needs to handle several words in a query (+15ms), therefore the search system can afford the absolute additional time.

\subsection{How to Ensure Robustness?}
Robustness proves to be a challenge in multiple tasks, since deep learning models are more likely to overfit the training data and over generalize word semantics, compared to traditional methods.  Our solution is to manipulate the training data and reuse handcrafted features to enforce robustness.
\begin{itemize}
    \item \textbf{Training data manipulation}. In query suggestion, we observe that with generalization pairs ("senior research scientist" -$>$ "research scientist"), the trained seq2seq will mostly generate queries by deleting original words.  We address it by removing the generalization pairs in the training data.
    \item \textbf{Reuse handcrafted features}. In document ranking, the deep NLP model may rank the topically related documents higher than the keywords matched documents. For example, in people search, matching professor profiles to query "student". This can be alleviated by incorporating the existing handcrafted features (\textit{e.g.}, keywords matching features such as cosine similarity) into neural networks.
\end{itemize}

\section{Related Work}
The task specific related work is mentioned in Section \ref{section:five-task}, therefore, in this section, we mainly focus on the common text based features and machine learning approaches.

\subsection{Traditional NLP Approaches}
For language understanding tasks, the text based features are usually sparse features such as unigrams, bigrams \cite{finkel2005incorporating} (classification and sequential tagging tasks), and dense matching features between queries and documents such as BM25 \cite{robertson:09} (ranking tasks).  In terms of algorithms, logistic regression, SVM, neural networks or decision tree \cite{cortes1995,zhang2000neural,burges2010} are used for classification and ranking.  The latter three introduce feature non-linearity but are usually not applied on sparse text features. For language generation tasks, usually there are two separate steps: candidate generation and candidate ranking \cite{bar2011,fonseca2005,cao2008,Shokouhi2013,li2006}. The generation is usually done by string matching, without explicitly modeling the word semantics. Ranking is usually similar to document ranking.

\subsection{Deep NLP Models}
CNN/LSTM/Transformer \cite{Lecun1995,vaswani2017,Hochreiter1997} are used to generate contextualized embeddings for words and sentences.  These embeddings are able to replace the sparse textual features such as unigrams/bigrams (in classification and sequential tagging) and textual matching features (in ranking). For language generation tasks, a language model with beam search decoding can perform the generation and ranking at the same time.  Attention \cite{bahdanau2014neural} is usually used in sequence-to-sequence framework \cite{Sutskever2014}. Recently, the pre-trained language models \cite{peters2018,radford2018,devlin2019} have shown impressive results on many NLP tasks by exploiting unsupervised data.

\subsection{Deep NLP for Search}
As mentioned in Section \ref{section:five-task}, deep NLP models have been applied in most of the search tasks \cite{hashemi2016,Huang2013,mitra2015,He2016,guo2016,xiong2017,dai2018}.
Additional information has been incorporated into the deep NLP models, such as personalization \cite{Jaech2018,fiorini2018}, session-aware \cite{Dehghani:2017,Ren2018}, etc.

Instead, the focus of this paper is to overcome the challenges of productionizing deep NLP models: latency, robustness, effectiveness \cite{mitra2018}.  All the resulting models have been deployed in the LinkedIn commercial search engines.  For example, an efficient BERT based ranking model is designed to enable document pre-computing which is crucial for productionization. In contrast, the existing BERT based ranking models  \cite{dai2019,macavaney2019,nogueira2019} do not allow document pre-computing.

There are some existing efforts that report online experiments, however, most models are for the document ranking task \cite{ramanath2018,yin2016,Grbovic2018,li2019}.  In this paper, we have a set of much broader tasks that covers query understanding and language generation.

\section{Conclusions}
Industry provides its own set of challenges, different in key ways from classic studied NLP tasks: (1) the amount and type of data, (2) constraints on latency or infrastructure, and (3) the highly optimized production baselines. This paper focused on how to apply deep NLP models to five representative search productions, illuminating the  challenges along the way.  All resulting models except query tagging are deployed in LinkedIn's search engines.   
More importantly, we also summarized the lessons learned across the tasks.  We listed the factors to consider when estimating the potential improvement brought by deep NLP models for a new task. We showed robustness and overfitting can be typically handled with careful data analysis. Latency, which is almost always a concern, can be addressed creatively in many ways, such as architecture simplification, two-pass systems, memorization, or parallel computation. 

\bibliographystyle{ACM-Reference-Format}
\bibliography{deepnlp}

\end{document}